\documentclass[a4paper,11pt]{article}
\usepackage{pos}

\title{Searching for the Origin of Double-peaked Broad Emission Lines in a Merging Galaxy with the EVN}
 \ShortTitle{EVN observation of a merging galaxy}

\author*[a,b,c]{Krisztina \'Eva Gab\'anyi}
\author[b,d]{S\'andor Frey}
\author[b]{Emma Kun}
\author[e]{Zsolt Paragi}
\author[f]{Tao An}

\renewcommand{\printHeadAuthors}{Gab\'anyi et al.}

\affiliation[a]{ELTE E\"otv\"os Lor\'and University, Institute of Geography and Earth Sciences, Department of Astronomy,\\
  P\'azm\'any P\'eter s\'et\'any 1/A, Budapest, Hungary}

\affiliation[b]{Konkoly Observatory, ELKH Research Centre of Astronomy and Earth Sciences,\\
Konkoly Thege Mikl\'os \'ut 15-17, Budapest, Hungary}

\affiliation[c]{ELKH-ELTE Extragalactic Astrophysics Research Group, ELTE E\"otv\"os Lor\'and University,\\
P\'azm\'any P\'eter s\'et\'any 1/A, Budapest, Hungary}

\affiliation[d]{Institute of Physics, ELTE E\"otv\"os Lor\'and University,\\
P\'azm\'any P\'eter s\'et\'any 1/A, Budapest, Hungary}

\affiliation[e]{Joint Institute for VLBI ERIC,\\
Oude Hoogeveensedijk 4, Dwingeloo, the Netherlands}

\affiliation[f]{Shanghai Astronomical Observatory, Key Laboratory of Radio Astronomy, Chinese Academy of Sciences,\\
80 Nandan Road, Shanghai, People's Republic of China}

\emailAdd{k.gabanyi@astro.elte.hu}
\emailAdd{frey.sandor@csfk.org}
\emailAdd{kun.emma@csfk.org}
\emailAdd{zparagi@jive.eu}
\emailAdd{antao@shao.ac.cn}

\abstract{The current cosmological structure formation models predict that galaxies evolve through frequent mergers. During these events, the supermassive black holes (SMBHs) residing in the centres of the galaxies shrink to the central region, while losing energy via dynamical friction. Detection of SMBHs in these galaxy mergers is straightforward if they are actively accreting matter from their surroundings as active galactic nuclei (AGN). Currently, only a few dual AGN are known. One way to identify dual AGN candidates is by detecting double-peaked emission lines in their spectra. If these are broad spectral lines, it may indicate the existence of two broad line regions associated with two AGN. 2MASS J165939.7$+$183436 is a merging system, where the detected double-peaked broad emission lines can be explained by a dual AGN with an estimated separation of $0.085''$. Radio emission from this object was detected in the Faint Images of the Radio Sky at Twenty-centimeters survey with a flux density of $\sim 2.6$\,mJy. We used the European Very Long Baseline Interferometer Network and the enhanced Multi-Element Remotely Linked Interferometer Network to image 2MASS J165939.7$+$183436 at $1.7$\,GHz. We did not detect compact radio emission from the source.}

\FullConference{%
  *** European VLBI Network Mini-Symposium and Users' Meeting (EVN2021) ***\\
  *** 12-14 July, 2021 ***\\
  *** Online ***
}

\begin{document}
\maketitle

\section{Introduction}

The current cosmological structure formation models predict that galaxies grow through frequent mergers (e.g., \cite{Volonteri2003}). During these events, the supermassive black holes (SMBHs) residing in the centres of the galaxies shrink to the central region of $\sim 1$ kpc and there, eventually, a bound SMBH pair may form \cite{1980Natur}. Detection of an SMBH is straightforward if it is actively accreting matter from its surroundings, as in active galactic nuclei (AGN). According to numerical simulations \cite{Capelo2015}, the simultaneous activity in the nuclei of merging galaxies are expected at a separation of $<10$ kpc. If the merging SMBHs have unequal mass or are asymmetric with respect to their spins, then the emitted gravitational waves are also asymmetric, which can result in a recoil \cite{Blecha2011} and may give rise to a centrally offset AGN. 

Currently, only a few dual AGN are known which could be spatially resolved, and thus their duality could be directly confirmed (e.g., \cite{An2018}). The most closely separated directly imaged binary AGN was discovered via serendipitous Very Long Baseline Interferometry (VLBI) observations \cite{Rodriguez2006}. VLBI offers the highest angular resolution, thus theoretically providing the best tool to resolve dual AGN. However, its usefulness is limited to a small fraction of potential sources, since only $\sim10$\% of AGN are radio-emitting \cite{Ivezic2002}.

Red AGN with colours of $J-K>1.7$, $R-K>4.0$, $E(B-V)>0.1$ are proposed \cite{Glikman2007} to be possible candidates of binary and/or recoiling AGN, since simulations showed that they could be explained as galaxy mergers in their final stage \cite{Blecha2011,Hopkins2008}. Double-peaked broad and narrow emission lines are also proposed to be signatures of binary AGN. However, the double peaks can also arise from the complex kinematics of the line-emitting regions (e.g., \cite{quest}), jet--cloud interaction (e.g., \cite{jetcloud}), and a rotating, disc-like broad line region (e.g., \cite{disclike}).  For the identification of recoiling AGN, spatial offset between the broad line region and the host galaxy (e.g., \cite{spatial_offset}), or a velocity offset of the broad lines with respect to the systemic velocity are proposed (e.g., \cite{vel_offset}).

2MASS\,J165939.7$+$183436 (hereafter J1659$+$1834) was identified as an AGN with double-peaked broad emission lines in a host galaxy located at a redshift $z=0.17$ and showing merger features \cite{Kim2020}. The extremely red colours of the AGN are attributed to dust obscuration. Optical and near-infrared spectra of J1659$+$1834 consistently show similar velocity offsets between the primary (blue) and secondary (red) broad emission line components. These components were detected for the hydrogen lines H$\alpha$, H$\beta$, Paschen$\beta$, and Paschen$\alpha$. If these lines are interpreted as arising from two different broad line regions (BLR), the H$\alpha$ lines imply SMBH masses of $10^{8.92\pm0.06}\mathrm{M}_\odot$ and $10^{7.13\pm0.06}\mathrm{M}_\odot$ \cite{GreeneHo}. Kim et al. \cite{Kim2020} estimated that the most plausible separation of the two line-emitting regions is $0.085''$, corresponding to $\sim 250$\,pc projected linear distance at the redshift of the source.

According to the Faint Images of the Radio Sky at Twenty-centimeters (FIRST, \cite{first}) survey, there is a radio source of $(2.56 \pm 0.15)$\,mJy flux density, at $\sim 2.2''$ to the southwest from the optical/infrared positions of J1659$+$1834. We observed this radio source with the European VLBI Network (EVN) $+$ the enhanced Multi-Element Remotely Linked Interferometer Network (e-MERLIN) to reveal whether the radio emission originates from AGN or star formation.

\section{Observation and Data Analysis}

The $1.7$-GHz VLBI observations of J1659$+$1834 took place on 2021 March 30 (project code: EG112). The array consisted of Jodrell Bank Mk2, Cambridge, Darnhall, Defford, Knockin, Pickmere (United Kingdom; the latter five from e-MERLIN), Westerbork (The Netherlands), Effelsberg (Germany), Medicina, Noto, Sardinia (Italy), Onsala (Sweden), Tianma (China), Toru\'n (Poland), Hartebeesthoek (South Africa), Svetloe, Zelenchukskaya, Badary (Russia), and Irbene (Latvia). All antennas except Irbene provided data. The observation was conducted in phase-reference mode, using the International Celestial Reference Frame (ICRF) source J165634.0$+$182626 as the phase-reference calibrator. The on-source integration time was $6$\,h. 
Except for the five e-MERLIN antennas, four intermediate frequencies (IFs) were used, each with a bandwidth of $32$\,MHz and each divided to $64$ channels.

For initial calibration, flagging and fringe fitting, we used the standard routines of the Astronomical Image Processing System (AIPS, \cite{aips}) software package. After a priori calibration and fringe-fitting of the phase-reference source, we imaged its structure with Difmap \cite{difmap} via the hybrid mapping method. The resulting model was used to determine the amplitude correction factors, which were then applied to the visibility amplitudes in AIPS. Additionally, the map of the phase-reference source was used as an input model for its subsequent fringe-fitting. The solutions were interpolated and applied to the target source, and the calibrated visibility data were then exported to be imaged in Difmap. 

\section{Results}

\begin{figure}
\centering
\includegraphics[bb=200 160 650 630, clip=, width=0.49\textwidth]{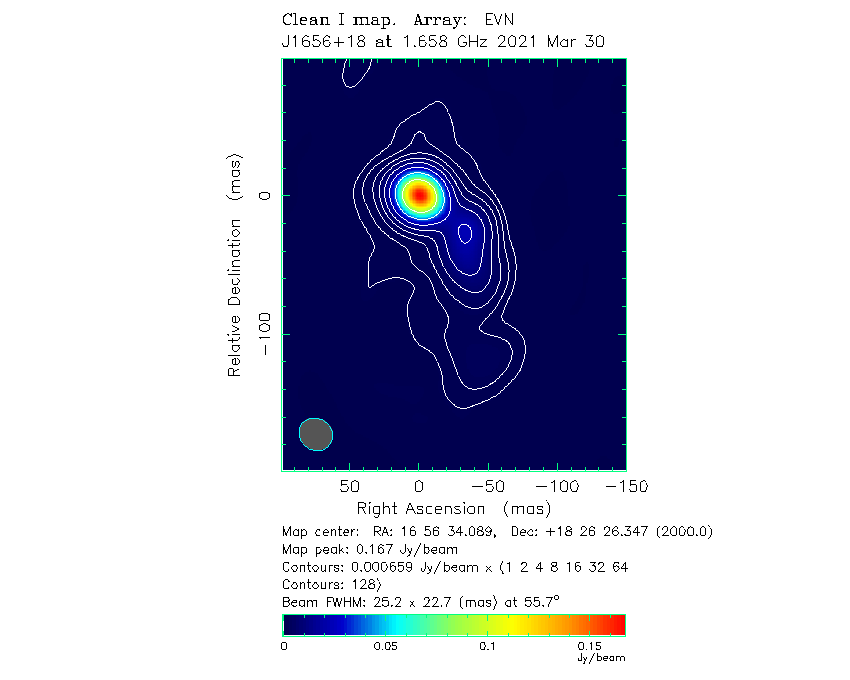}
\includegraphics[bb=200 160 650 630, clip=, width=0.49\textwidth]{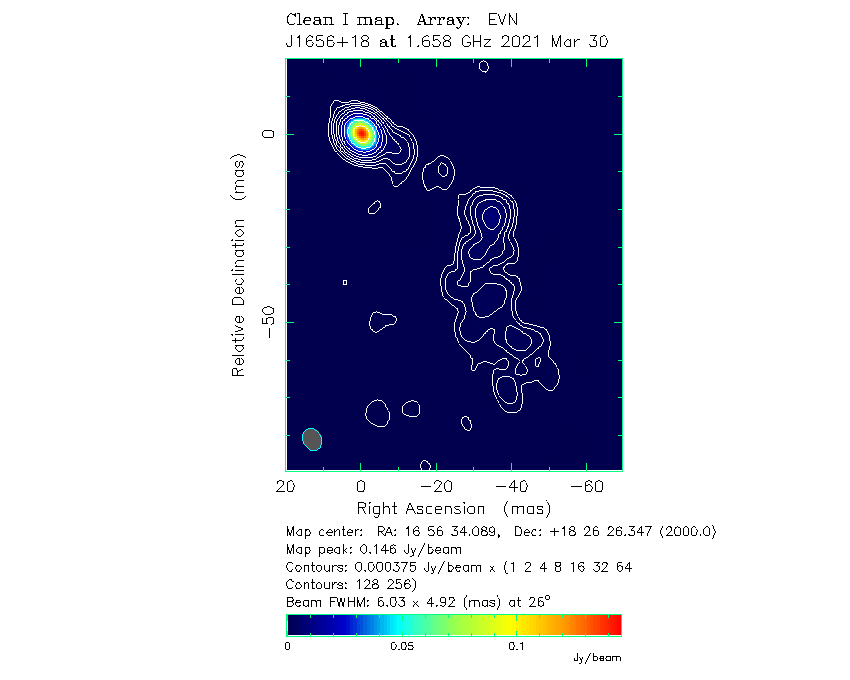}
\vspace{-0.3cm}
\caption{$1.7$-GHz EVN+e-MERLIN maps of the phase-reference calibrator source, J165634.0$+$182626. {\it Left: }The long baselines are down weighted by a Gaussian taper of $0.5$ at $5$\,M$\lambda$ radius. The peak intensity is $167$\,mJy\,beam$^{-1}$. The lowest contour is $0.7$\,mJy\,beam$^{-1}$, further contours increase by a factor of $2$. The restoring beam is shown at the lower left corner of the image, its full-width half maximum size is $25.2$\,mas $\times 22.7$\,mas with a major axis position angle of $55.7^{\circ}$. {\it Right:} Naturally weighted image. The peak intensity is $146$\,mJy\,beam$^{-1}$. The lowest contour is drawn at $6\sigma$ image noise level, $0.4$\,mJy\,beam$^{-1}$, further contours increase by a factor of $2$. The restoring beam is shown at the lower left corner of the image, its full-width half maximum size is $6.03$\,mas $\times 4.92$\,mas with a major axis position angle of $26.0^{\circ}$.\label{fig:cal}}
\end{figure}

We did not detect radio emission at the position of J1659$+$1834 at $1.7$\,GHz, down to a $6\sigma$ image noise level of $50\,\mu$Jy\,beam$^{-1}$. Thus, we cannot confirm the existence of a radio-emitting AGN in the system. However, we cannot rule out its existence, either, since extended emission from the lobe(s) of radio-loud AGN could have been resolved out in our interferometric observation. In that case, since the assumed jet axis is close to the plane of the sky, no boosted compact emission is expected from the central part, and thus the possibly very radio-faint compact central region cannot be detected. The extended radio emission detected by FIRST thus, can arise from star formation, enhanced due to the merger event, or by large-scale emission from the lobe(s) of a radio-AGN, or from the combination of both. If the FIRST-detected radio emission solely comes from star formation, it implies a star formation rate of a moderate $\sim 11\, \mathrm{M}_\odot\mathrm{yr}^{-1}$ according to the scaling relation of \cite{Hopkins2003}. Thus, star formation alone can explain the radio emission making  the contribution from a resolved lobe emission scenario unnecessary. In that case, a non-radio emitting AGN still exists in the system, and the peculiar broad emission lines can be explained as arising from a rotating, disc-like broad line region. A more sensitive, low-frequency VLBI radio observation can image the structure of the FIRST-detected radio emission and may disentangle star formation and AGN-related emission.

We also imaged the phase-reference calibrator (J165634.0$+$182626) in Difmap after the improved fringe-fitting. We present two maps of the source in Fig.\,\ref{fig:cal}. In the left panel, the long baselines are down weighted by using a Gaussian taper with a value of $0.5$ at $5$\,M$\lambda$. In the right panel, we show the naturally-weighted image. These maps reveal an extended, distorted low surface brightness jet in the south-southwest direction. To our knowledge, this is the first $1.7$-GHz VLBI image published for this high-redshift ($z=2.55$, \cite{SDSS}) quasar. VLBI observations conducted at $2.3$, $4.3$, $4.8$, $7.6$, $8.3$, $8.4$, and $8.7$\,GHz were available from the astrogeo.org website (maintained by L. Petrov) at the time of preparing our observations. These maps showed the compact bright core region of the source, enabling us to select J165634.0$+$182626 as a phase-reference calibrator. The most recent archival data taken at the lowest frequency ($2.3$\,GHz) where the observations are the most sensitive to steep-spectrum, extended emission indicated a feature oriented roughly $40$\,mas south-west from the compact core. Our EVN+e-MERLIN observation (Fig.\,\ref{fig:cal}) revealed in more detail this $\sim100$-mas long, bending complex jet structure.

\acknowledgments
The European VLBI Network is a joint facility of independent European, African, Asian, and North American radio astronomy institutes. Scientific results from data presented in this publication are derived from the following EVN project code(s): EG112. The e-MERLIN is a National Facility operated by the University of Manchester at Jodrell Bank Observatory on behalf of STFC. We thank the Hungarian National Research, Development and Innovation Office (OTKA K134213, 2018-2.1.14-T\'ET-CN-2018-00001) for support.

\end{document}